\documentclass{llncs}
\usepackage[sectionbib,authoryear]{natbib}
\usepackage{graphicx}

\begin{document}
\title{Probabilistic Approaches to Alignment with Tandem Repeats}
\author{Michal N\'an\'asi \and Tom\'a\v{s} Vina\v{r} \and Bro\v{n}a Brejov\'a}
\institute{Faculty of Mathematics, Physics, and Informatics,
           Comenius University,\\
           Mlynsk\'a dolina, 842 48 Bratislava, Slovakia
}
\maketitle

\begin{abstract}
We propose a simple tractable pair hidden Markov model for pairwise
sequence alignment that accounts for the presence of short tandem
repeats. Using the framework of gain functions, we design several
optimization criteria for decoding this model and describe the
resulting decoding algorithms, ranging from the traditional Viterbi
and posterior decoding to block-based decoding algorithms specialized
for our model. We compare 
the accuracy of individual decoding algorithms on simulated data and
find our approach superior to the classical three-state pair HMM 
in simulations.
\end{abstract}

\section{Introduction}

In this paper, we explore the use of pair hidden Markov models (pair
HMMs, PHMMs) in improving the quality of pairwise sequence alignment
in the presence of tandem
repeats. We propose a simple tractable model that explicitly accounts
for short tandem repeats, and we use the framework of maximum expected 
gain to explore a variety of decoding optimization criteria for our model.

Pair HMMs have for a long time played a
major role in sequence alignment \citep{Durbin1998}. 
The traditional Needleman-Wunsch
algorithm \citep{Needleman1970} and its variants can be easily
formulated as a special case of alignment with PHMMs (we call this
approach Viterbi decoding).  
The main advantage of PHMMs is that they allow to express the
scoring scheme in a principled way in the context
of a probabilistic model.

Sequence alignments are a mainstay of comparative
genomics. By comparing sequences that evolved from a common ancestor,
we can infer their phylogenetic relationships, discover sites under
functional constraint, or even shed light on the function of individual
sequence elements. However, comparative genomic methods are very
sensitive to the quality of underlying alignments, and even slight
inaccuracies may lead to artifacts in the results of comparative methods.

It is very difficult to evaluate alignment accuracy,
yet even simple statistics can reveal artifacts of present-day
algorithms. \citet{Lunter2008} demonstrated systematic biases caused
by the optimization criteria set by the Needleman-Wunsch approach. 
They show that by using
variants of the posterior decoding instead of the traditional
Viterbi algorithm, one can significantly increase the
quality of alignments. While the Viterbi decoding seeks one
highest scoring alignment, the posterior decoding summarizes
information from all alignments of the two sequences.
This approach was also found superior by other authors 
\citep{Miyazawa1995,Holmes1998,Schwartz2007}.

An algorithm by \citet{Hudek2010} is an intermediate between 
the Viterbi and posterior decoding, summarizing probabilities of alignments
within short blocks. 
The goal is to segment the alignment into blocks, where each block
has gaps in only one of the two sequences. The decoding algorithm
considers each block as a unit, summing probabilities of all
alignments that had the same block structure. Finally,
\citet{Satija2010} have demonstrated that fixing a particular
alignment is not necessary in some comparative genomics applications,
instead one can consider all possible alignments weighted by their 
probability in the PHMM.

In this paper, we concentrate on modeling sequence alignments in the
presence of tandem repeats. Short tandem repeats cover more than 2\%
of the human genome, and occur in many genes and regulatory regions
\citep{Gemayel2010}; in fact, majority of recent short insertions in
human are due to tandem duplication \citep{Messer2007}. 
Evolution of tandem repeats is dominated by tandem segmental
duplications resulting in regions composed of a
highly variable number of almost exact copies of a short segment. 
Such sequences are difficult to align with standard scoring schemes,
because it is not clear which copies from the two organisms are
orthologous. Misalignments due to the presence of short
tandem repeats are usually not limited to the repetitive sequence
itself, but may spread into neighbouring areas and impact the
overall alignment quality.


Sequence alignment with tandem duplication was first studied by
\citet{Benson1997}. They propose an extension of the traditional
Needleman-Wunsch algorithm that can accommodate tandem repeats in $O(n^4)$
time. They also propose several faster heuristic algorithms.
Additional work in this area concentrated on computing variants of 
edit distance either on whole sequences with tandem arrays or 
on two tandem arrays using different sets of evolutionary operations
\citep{Sammeth2006,Berard2006,Freschi2012}.

The first probabilistic approach to alignment of tandem duplications was
introduced by \citet{Hickey2011}, who developed a new probabilistic
model by combining PHMMs with Tree Adjoining Grammars (TAGs).  Their
model favors tandem duplications to other insertions, but the approach
does not explicitly model whole arrays of tandemly repeated motifs.
Moreover, algorithms to train and decode such
models are relatively complex. 

Protein sequences with repetitive motifs (such as zinc finger
proteins) are a special class of proteins and their alignment has many
features in common with DNA sequence alignment with tandem
repeats. \cite{Kovac2012} combined profile HMMs (capturing the
properties of the repeating motif) and PHMMs (modeling alignments)
into a single scoring scheme that can be decoded by a newly proposed
algorithm. However, their scoring scheme is no longer a probabilistic
model and the method is focused on correctly aligning individual
occurrences of a single motif rather than alignment of long sequences
interspersed with multiple motifs.

Here, we propose a simple tractable PHMM for sequence alignment
with tandem repeats, and we explore various decoding methods for use of
this model in sequence alignment. In addition to the classical Viterbi
decoding, we define several variants based on the posterior decoding and
block-based methods tailored to the specifics of our model.
To demonstrate the differences, we have implemented several
of these methods and compared their performance.


\section{Pair HMM for Alignment with Tandem Repeats}

Tandem repeats may arise by a complicated sequence of evolutionary
events, including multiple rounds of tandem duplication, deletion,
point mutation, gene conversion and other phenomena. Tandem repeat
arrays at homologous locations in two related species may have arisen
in the common ancestor and thus share part of their evolutionary
history, but they could be further modified by independent events
occurring after speciation. Models attempting to capture such diverse
evolutionary mechanisms usually lead to complex problems in inference
and parameter estimation. We propose a tractable model,
based on classical PHMMs, 
which still captures the essence of a tandem repeat array:
periodically repeating motif, which may be shared between the two
species, or be specific for one species only.

A PHMM defines a probability distribution over alignments of two
sequences $X$ and $Y$. The standard PHMM has three states: match
state $M$ generating ungapped columns of the alignment,
and two insert states $I_X$ and $I_Y$, where $I_X$
generates alignment columns with a symbol from $X$ aligned to a gap,
and $I_Y$ generates columns with a symbol from $Y$
aligned with a gap \citep{Durbin1998}. In our work, we will use a more
complex PHMM, but standard algorithms for inference in these models
are still applicable.

\begin{figure}
\includegraphics[width=5cm]{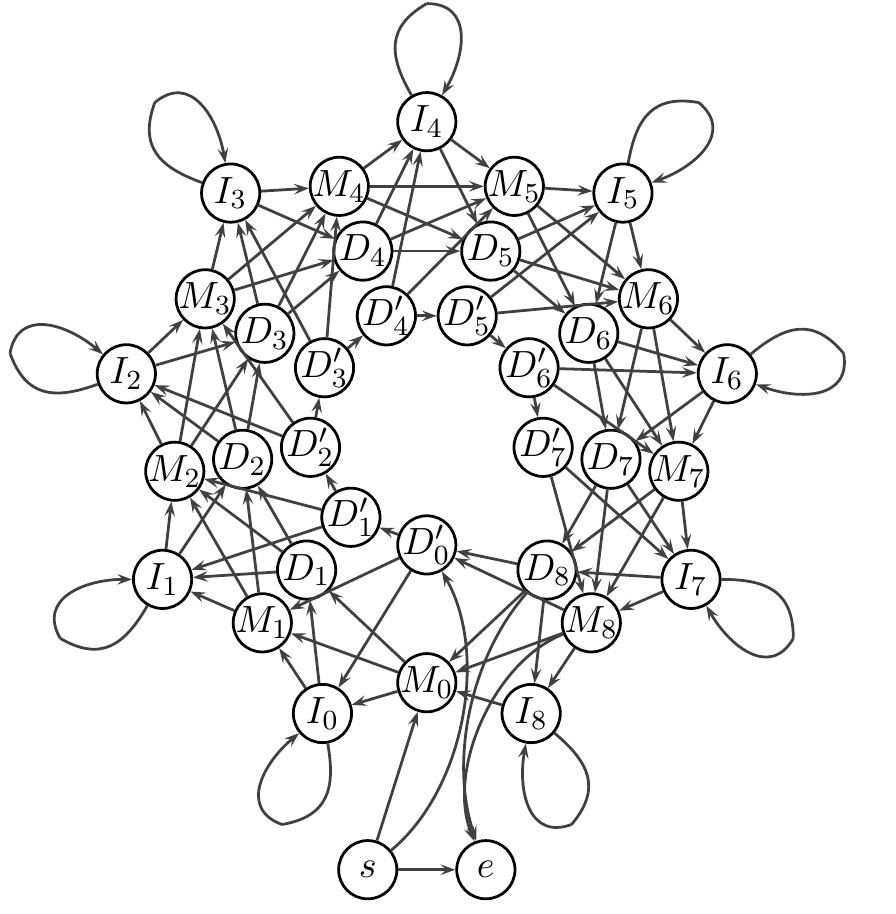}
\hfill
\includegraphics[width=5.5cm]{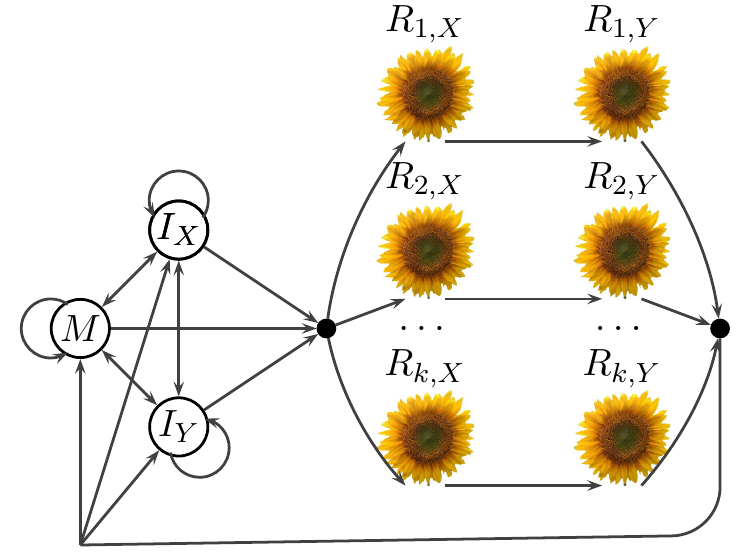}\\
\caption{SFF (sunflower field) model: a pair hidden Markov model for
  alignment with tandem repeats. Each submodel $R_{i,\alpha}$ (left)
  is a circular profile hidden Markov model emitting tandem copies of
  the motif in one sequence. State $M_j$ is the match state generating
  $j$th symbol of the motif, state $I_j$ allows insertions between
  symbols $j$ and $j+1$ of the motif, and states $D_j$ and $D'_j$
  allow to skip state $M_j$. States $s$ and $e$ designate the entry
  and exit points from the submodel. The full SFF model (right) contains a
  standard three-state PHMM with states $M$, $I_X$ and $I_Y$, and
  two submodels $R_{i,X}$, $R_{i,Y}$ for each motif $i$. States and
  submodels with subscript $X$ and $Y$ generate symbols in the
  respective sequence $X$ or $Y$ only.}
\label{fig:sff}
\end{figure}

We call our model SFF and its details are shown in
Fig.\ref{fig:sff}. The model contains a standard three-state PHMM
and two ``sunflower'' submodels $R_{i,X}$ and $R_{i,Y}$ for each
possible repeating motif $i$.  Submodel $R_{i,X}$ generates several (possibly
zero) copies of the motif in sequence $X$ and submodel $R_{i,Y}$
generates motif copies in sequence $Y$. Each copy of the
motif is generated independently and the number of copies in $X$
and $Y$ are independent and geometrically distributed.

Each sunflower submodel is a circularized profile HMM emitting copies of the
motif in one of the two sequences. For a motif of length $p$, the
submodel contains $p$ match states $M_0,\dots, M_{p-1}$, each match
state emitting one symbol of the motif. Insertion state $I_j$ allows
to insert additional characters between symbols emitted by $M_j$ and
$M_{(j+1)\bmod p}$. Deletion states $D_j$ and $D'_j$ allow to bypass
match state $M_j$, and thus correspond to deletions with respect to
the reference motif sequence. Since the submodel can emit multiple
tandem copies of the motif, the states in column $p-1$ are connected to
the states in column 0. To avoid cycles consisting solely of silent
states, we use two separate chains of deletion states. Chain
$D'_0,\dots, D'_{p-2}$ can be entered only in state $D'_0$, and model
can stay in this chain for at most $p-1$ steps. Chain $D_1,\dots, D_{p-1}$
can be entered only after visiting at least one match or insert state
in the current copy of the motif. As a result, the model can never
pass around the whole circle using delete states. Note that the model
prefers integer number of repeats, even though partial repeat occurrences
are common in the real data. If desired, this can be addressed by simple
changes in the model topology or parameters.

The overall model can have sunflower submodels for an arbitrary number
of motifs; we can even define an infinite model, in which every
possible finite string serves as the consensus for one pair of
sunflowers. In our work, we use $k=310,091$ motifs chosen as consensus strings
of all tandem repeats found by the TRF program \citep{Benson1999} run on the
human chromosome $15$ and its homologous sequences in the dog genome. 
The probability of choosing a particular motif
out of all $k$ possibilities can be uniform or dependent on the motif
length or composition. We assign this probability based on the observed
frequency of the corresponding consensus pattern in the TRF output.
Alternatively, we could use a much smaller model by \citet{Frith2011};
however, this model does not easily handle insertions and deletions 
within repeats.

Likewise, we could use a multiple alignment of real motif occurrences
to set individual parameters of the profile HMM. Instead, we use the
same set of parameters for all states of all motif submodels.  In
particular, we set the insert and delete rates to $0.005$; the match
states allow mutations away from consensus according to the
Jukes-Cantor model with parameter $t=0.05$. Parameters of the
three-state PHMM were estimated from the UCSC alignment of the human
chromosome $15$ and its homologous regions in the dog genome.

Our model also assumes that individual copies of a fixed motif are
independent. If they share part of their evolutionary history, this
assumption is not valid, but it greatly simplifies the model. We could
add some limited dependence by introducing repeat submodels
emitting copies in the two sequences simultaneously; we have used such
a model in a different setting in our previous work \citep{Kovac2012}.

\section{Inference Criteria and Algorithms}

Given the SFF model introduced in the previous section, and two
sequences $X=x_1\dots x_n$ and $Y=y_1\dots y_m$, 
we wish to find the alignment of these two
sequences best agreeing with the model. We can also annotate this
alignment by labeling individual alignment columns with additional
information. We start by defining an alignment and its annotation more
formally (see Fig.\ref{fig:alignment}). 
\emph{An alignment} of $X$ and $Y$ is a sequence of pairs
$(a_1,b_1),\dots (a_t,b_t)$, each pair representing one alignment
column. Symbol $a_i$ represents either a position in $X$, or a gap
annotated with the position of the nearest non-gap symbol on the left;
formally $a_i\in \{1,\dots,n\} \cup \{-_0,-_1,\dots,-_n\}$. To specify
a valid alignment, $a_1$ must be $1$ or $-_0$, $a_t$ must be $n$ or
$-_n$, and if $a_i\in \{j, -_j\}$, $a_{i+1}$ must be $j+1$ or
$-_j$. The conditions on symbols $b_i$ representing positions in
sequence $Y$ are analogous. The \emph{state annotation} of an alignment is a
sequence of states $s_1\dots s_t$ such that state $s_i$ generated
alignment column $(a_i,b_i)$. The \emph{repeat annotation} 
is a binary sequence $r_1\dots r_t$, where $r_i=1$ if the
state $s_i$ generating the $i$-th column is one of the states in the repeat
submodels. While the state annotation can be used with any PHMM
generating the alignment, the repeat annotation is appropriate only for the 
SFF model or other PHMMs explicitly modeling repeats.

\begin{figure}[t]
\begin{tabular}[t]{@{}c*{7}{@{\quad}c}@{}}
$X$: & A & C & - & - & - & - & T \\
$Y$: & - & C & G & A & A & A & T \\
\end{tabular}
\hfill
\def\M#1{$-_{#1}$}
\def\R{$r$}
\begin{tabular}[t]{@{}c*{7}{@{\quad}c}@{}}
$a_i$: &  1  &  2  & \M2 & \M2 & \M2 & \M2 &  3 \\
$b_i$: & \M0 &  1  &  2  &  3  &  4  &  5  &  6 \\
$s_i$: &$I_X$& $M$ &$I_Y$& \R  &  \R  &  \R  &  R \\
$r_i$: &  0  &  0  &  0  &  1  &  1  &  1  &  0 \\
\end{tabular}
\caption{Example of an alignment represented in our notation, together
  with its state and repeat annotation. Assuming that submodel $R_{1,Y}$ 
in the SFF model represents consensus sequence {\tt A}, 
state $r$ in the state sequence 
is a shorthand for the state $M_1$ within $R_{1,Y}$.
}
\label{fig:alignment}
\end{figure}

We will explore several inference
criteria for choosing the best alignment. To describe them, we will use
the terminology of gain functions \citep{Hamada2009}; analogous
notion of a loss functions is frequently used for example in
statistics and machine learning.  A {\em gain function} $G(A,A_T)$
evaluates similarity between a predicted alignment $A$ and the correct
alignment $A_T$; higher gain meaning that the prediction is of higher
quality. Since the true alignment $A_T$ is not known, we will consider
the expected gain $E_{A_T}[G(A,A_T)|X,Y]$ of alignment $A$, assuming that
sequences $X$ and $Y$ were generated by our model
$$E_{A_T}[G(A|A_T)|X,Y] = \sum_{A_T} G(A,A_T)\Pr(A_T|X,Y).$$
In each optimization criterion, we choose a particular gain function
and look for alignment $A^*$ maximizing the expected gain
$A^* = \arg\max_A E_{A_T}[G(A,A_T)|X,Y]$.
Note that
the gain function is only a way of defining the optimal solution; the
corresponding decoding algorithm needs to be designed on a case-by-case
basis.  

\subsection{Decoding Criteria for the Three-State PHMM}
For simplicity, we start with criteria for the three-state PHMM, where
the state annotation is uniquely determined by the alignment itself. 

\paragraph{Viterbi decoding.} Perhaps the simplest 
gain function assigns gain +1 if the predicted alignment $A$ is
identical to the true alignment $A_T$, and 0 otherwise. To optimize
this gain function, we need to find the alignment with the highest
overall probability in the model. In the simple three-state PHMM, this
alignment can be found by the classical Viterbi algorithm in time
$O(nmE)$, where $E$ is the number of non-zero transitions in the model.

\paragraph{Posterior decoding.} 
While the Viterbi decoding assigns gain only if the whole alignment is
correctly predicted, posterior decoding assigns gain +1 for each
correctly identified alignment column. Recall that the column is a
pair $(a_i, b_i)$, and it is considered correct, if the same column
also occurs somewhere in the true alignment. The optimal 
alignment under this gain function can be
found by computing the posterior probability of each alignment column
using the forward and backward algorithms for PHMMs, and then finding 
the alignment as a collection of compatible columns with the highest
sum of posterior probabilities. A similar algorithm is considered for
example by \citet{Lunter2008}, except that the column posteriors are 
multiplied rather than added. The running time of this algorithm is
again $O(nmE)$.

\paragraph{Marginalized posterior decoding.}
\citet{Lunter2008} also consider a variant of posterior decoding,
where a column $(i,-_j)$ is considered correct and receives a gain +1,
if the true alignment contains a column $(i,-_{\ell})$ for any value
of $\ell$. In other words, when
symbol $x_i$ is aligned to a gap, we do not distinguish where
is the location of this gap with respect to sequence $Y$.
Columns $(-_j,i)$ are treated symmetrically.
To optimize this gain function, we again start by computing posteriors
of all columns. Then we marginalize the probabilities of gap
columns, effectively replacing posterior of column $(i,-_j)$ 
with the sum of posteriors of columns $(i,-_\ell)$ for all $\ell$. 
As before, we then find the alignment maximizing the sum of posteriors
of its columns. The algorithm runs in $O(nmE)$ time. 

\subsection{Decoding Criteria for the SFF Model}

In more complex models, including ours, one alignment can be generated
by several different state paths. Various gain functions can thus take
into account also the state or repeat annotation of the alignment.

\paragraph{Viterbi decoding.} 
In more complex models, the classical Viterbi algorithm optimizes 
a gain function in which the alignment is
annotated with the state path generating it, and gain is awarded only
when both the alignment and the state path are completely
correct. 

\paragraph{Posterior and marginalized posterior decoding.} 
We will consider a variant of the posterior decoding, in which alignment
columns are annotated by the repeat annotation, and an alignment
column gets a gain +1, if the true alignment contains the same column
with the same label. The only change in the algorithm is that 
the forward-backward algorithm produces posterior probabilities of
columns annotated with the state, which are then marginalized over all
states with the same repeat label. The running time is still $O(nmE)$. 
Similar modification can be done for marginalized posterior decoding,
where we marginalize gap columns based on both state and gap position.

\paragraph{Block decoding.}
We will consider also a stricter gain function, which requires that
repeat regions have correctly identified boundaries. We will split the
alignment annotated with repeats into \emph{blocks}, so that each 
maximal region of
consecutive columns labeled as a repeat forms a block.  Each column
annotated as a non-repeat also forms a separate block.  The gain
function awards gain +1 for each non-gap symbol in every correctly
predicted and labeled block. Correctness of non-repeat columns is
defined as in posterior decoding. A repeat block is
considered correct, if exactly the same region in $X$ and the same
region in $Y$ are also forming a repeat block in the true alignment.
Note that the gain for each block is proportional to the number of
non-gap symbols in the block to avoid biasing the algorithm towards
predicting many short blocks. 

To optimize this gain function, we first compute posterior
probabilities for all blocks. Note that a block is given by a pair of
intervals, one in $X$ and one in $Y$. Therefore the number of 
blocks is $O(n^2m^2)$. The expected gain of a block is its posterior
probability multiplied by the number of its non-gap symbols.
After computing expected gains of individual blocks, we can find
the highest scoring combination of blocks by dynamic programming in
$O(n^2m^2)$ time.

To compute block posterior probabilities, we transform the SFF model
to a generalized PHMM \citep{Pachter2002},
in which all repeat states are replaced by a
single generalized state $R$. In generalized HMMs, emission of a state
in one step can be an arbitrary string, rather than a single character. In
our case, the new state $R$ generates a pair of sequences from 
the same distribution as defined by one pass through the repeat
portion of the original SFF model. Pair of sequences generated by $R$ 
represents one block of the resulting alignment. We call this new
model the block model. Using the forward-backward algorithm for
generalized HMMs, we can compute posterior probabilities of all blocks in 
$O(n^2m^2f)$ time where $f$ is the time necessary to compute emission
probability for one particular block. 

If we naively compute each emission separately, we get $f=O(nmE)$. 
However, we can reduce this time as follows. If the SFF contains only
one motif, the emission probability of sequences $x$ and $y$ in the $R$
model is simply 
\[\Pr\left(x,y\mid
R\right)=\Pr\left(x\mid R_{1,X}\right)\Pr\left(y\mid R_{1,Y}\right),\]
because the model first generates $x$ in the sunflower submodel $R_{1,X}$ and
then generates $y$ in the model $R_{1,Y}$. Note that these two models
are connected by a transition with probability $1$. 
In the general case, we sum the probabilities 
for all $k$ motifs, each multiplied by the transition probability of
entering that motif. To compute block emission probabilities fast, 
we precompute $\Pr\left(x\mid R_{i,X}\right)$ and $\Pr\left(y\mid
R_{i,Y}\right)$ for all substrings $x$ and $y$ of sequences $X$ and
$Y$ respectively. This can be done by the forward algorithm 
in $O((n^2 + m^2)E)$ time. After this preprocessing, the
computation of emission probability is $O(k)$, and the overall
running time of this algorithm is $O(kn^2m^2 + (n^2 + m^2)E)$.

%
%

\paragraph{Block Viterbi decoding.}
The final gain function we consider is a variant of the Viterbi
decoding. The Viterbi decoding assigns gain +1 for a completely
correct alignment labeled with a correct state annotation. 
One alternative is to assign gain +1 if the alignment and its repeat
annotation are completely correct. This gain function considers 
as equivalent all state paths that have the same position of repeat
boundaries but use different motifs or different alignments of the
sequence to the motif profile HMM. 

In the SFF model, location of a repeat block uniquely
specifies alignment within the block, because all symbols from
sequence $X$ must come first (aligned to gaps), followed by symbols
from sequence $Y$. However, some models may emit repeat bases from the
two sequences aligned to each other.  We wish to abstract from exact
details of repeat alignment, and consider different alignments within
a repeat as equivalent. Therefore, we will reformulate the gain
function in terms of blocks. The alignment labeled with repeat
annotation gets a gain 1, if all its blocks are correct, where block
correctness is determined as in the block decoding.
This formulation is similar to the one solved by \citet{Hudek2010}.

To optimize this gain function, we use the Viterbi algorithm for
generalized HMMs applied to the block model, which leads to running
time $O(kn^2m^2 + (n^2 + m^2)E)$, by similar reasoning as above. 

\subsection{Practical considerations}

Even the fastest algorithms described above require $O(nmE)$ time,
where sequence lengths $n$ and $m$ can be quite high when aligning
whole syntenic genomic regions and the size of the model $E$ depends on
the sum of the lengths of all repeat motifs, which can be potentially
even infinite. However, we can use several heuristic approaches to
make the running times reasonable. 

First of all, we can use the standard technique of banding, where we
restrict the alignment to some window around a guide alignment
obtained by a faster algorithm. A simpler form of banding is to split
the guide alignment to non-overlapping windows and realign each window
separately. These techniques reduce the $O(nm)$ factor. 

To restrict the size of the model, we first find tandem repeats in 
$X$ and $Y$ independently by running the TRF program
\citep{Benson1999}. Then we include in our model only those motifs
which appear at least once in the TRF output. If we process only
relatively short windows of the banded alignment, the size of the
model will be quite small. Note however, that we keep the transition
probabilities entering these models the same as they are in the full
SFF model. If TRF finds a consensus not included in the original SFF model, 
we add its two submodels with a small probability comparable to the
rarest included motifs.

These two heuristics sufficiently speed up algorithms running in
$O(nmE)$ time. The block decoding and the block Viterbi decoding need
to consider all possible blocks, which is prohibitive even within
short alignment windows.  Therefore, we limit
possible repeat blocks only to intervals discovered as repeats by the TRF
program. We allow the generalized repeat state $R$ to 
generate the block of substrings $x$ and $y$ if each of these
substrings is either empty or one of the intervals found by TRF 
has both its endpoints within $10$ bases from the respective endpoints
of $x$ or $y$. Therefore, if TRF finds $t_X$ intervals in $X$ and
$t_Y$ intervals in $Y$, we try at most $(20 t_X + n)(20t_Y +m)$ blocks.

The final consideration is that the SFF model does not align tandem
repeats at orthologous locations, even if they share a common
evolutionary history. This might be impractical for further use. 
Therefore we postprocess the alignments by realigning all
blocks annotated as repeats using the standard three-state PHMM.
In this realignment, we also include gaps adjacent to these repeats.

\section{Experiments}

We have compared decoding methods described in the previous section
and several baseline algorithms on simulated data (see Table
\ref{simulated_experiment}). The data set contained 200 alignments of
length at least 200 each generated from the SFF model (the same model
parameters were used in the sampling and for the alignments). In
generating the dataset, we required that each tandem repeat had at
least three copies in both species; otherwise, we would obtain many
regions that would be labeled as tandem repeats, but would in fact
only have a single copy. The error rate (the first column of the table)
measures the fraction of true alignment columns that were not
found by a particular algorithm. It
was measured only on the alignment columns that were generated from 
non-repeat states in the simulation, as the SFF model 
does not give any alignment in repeat regions.

\begin{table}[t]
\def\M{$^*$} 
\def\MM{$^{**}$} 
\caption{Accuracy of several decoding methods on simulated data.
\M: method uses the real consensus motifs. \MM: method uses 
the real consensus motifs and intervals from the real repeat blocks.
}
\label{tab:sim}
\def\CC#1{\multicolumn{1}{c}{#1}} 
\centerline{\begin{tabular}{lr@{\quad}rr@{\quad}rr}
\hline
          & \CC{Alignment} & \multicolumn{2}{c}{Repeat} & 
\multicolumn{2}{c}{Block}\\
Algorithm & \CC{error} & \CC{sn.} & \CC{sp.} & \CC{sn.} & \CC{sp.} \\
\hline
\hline
SFF marginalized    & {\bf 3.37\%} & {\bf 95.97\%} & 97.78\% & {\bf 43.07\%} & 44.87\%\\
SFF posterior       & 3.53\% & 95.86\% & 97.87\% & 42.70\% & 47.37\%\\
SFF block           & 3.87\% & 91.20\% & {\bf 98.04\%} & 36.13\% & 47.14\%\\
SFF block Viterbi   & 4.32\% & 91.28\% & 97.96\% & 35.40\% & 45.97\%\\
SFF Viterbi         & 4.04\% & 95.29\% & 97.85\% & 42.70\% & {\bf 48.95\%}\\
\hline
SFF marginalized\M  & 3.02\% & 98.93\% & 99.64\% & 77.01\% & 76.17\% \\ 
SFF posterior\M     & 3.42\% & 98.84\% & 99.51\% & 75.91\% & 80.93\% \\
SFF block\MM        & 3.21\% & 97.70\% & 99.87\% & 80.66\% & 94.44\% \\
SFF block Viterbi\MM& 3.71\% & 98.12\% & 99.85\% & 81.75\% & 92.18\% \\
SFF Viterbi\M       & 3.94\% & 98.54\% & 99.45\% & 75.55\% & 83.47\% \\
\hline
Context             & 5.98\% \\
Muscle              & 5.62\% \\
3-state posterior   & 4.41\% \\
3-state Viterbi     & 4.78\% \\
\hline
\end{tabular}}
\label{simulated_experiment}
\end{table}

The first observation is that the methods based on the SFF model
(the first block of the table) outperform the baseline method (the Viterbi
algorithm on the three-state model), reducing the error rate by 10--30\%.
In general, the methods that score individual alignment 
columns are more accurate than the
block-based or the Viterbi-based methods, which is not surprising, because
error rate as a measure of accuracy is closer to the gain function they
optimize. We have also compared our approach to the related 
method of context-sensitive indels \citep{Hickey2011}.
The Context program was trained on a separate set of 200 alignments
sampled from our model. However, its error rate is quite high, 
perhaps due to insufficient training data or some software issues.
Finally, we have also run the Muscle aligner with default parameters
\citep{Edgar2004}; we have used the result as a guide alignment
for the slower block-based decoding methods (the new alignment was
restricted to be within 30 base window from the guide
alignment). 

The SFF-based algorithms use the
tandem repeat motifs predicted by the TRF, as well as approximate 
repeat intervals (block-based methods). The TRF predictions are not
exact and may contribute to the overall error rate. 
We attempted to quantify this effect by using
the real tandem repeat motifs and real boundary positions instead of
the TRF predictions (the second block of Table \ref{simulated_experiment}). 
We can see that the use
of TRF predictions indeed leaves space for improvement,
with the best performing algorithm reducing the error rate by almost
40\% compared to the baseline. Block-based methods work significantly
better with true intervals than with the TRF intervals, suggesting
that further improvements in repeat interval detection are needed.

The decoding methods that use the SFF model produce an alignment and a
repeat annotation. Comparing annotation of each base in both sequences
with the true repeat annotation sampled from the model (table columns
repeat sensitivity and specificity), we note that the marginalized
posterior decoding is the most sensitive, and the block decoding the
most specific method. Specificity was quite high for all methods, low
sensitivity for block-based methods was probably caused by wrong
repeat intervals predicted by the TRF, since it improves markedly by
using correct intervals. 

\begin{figure}[t]
\centerline{\includegraphics[width=0.6\textwidth]{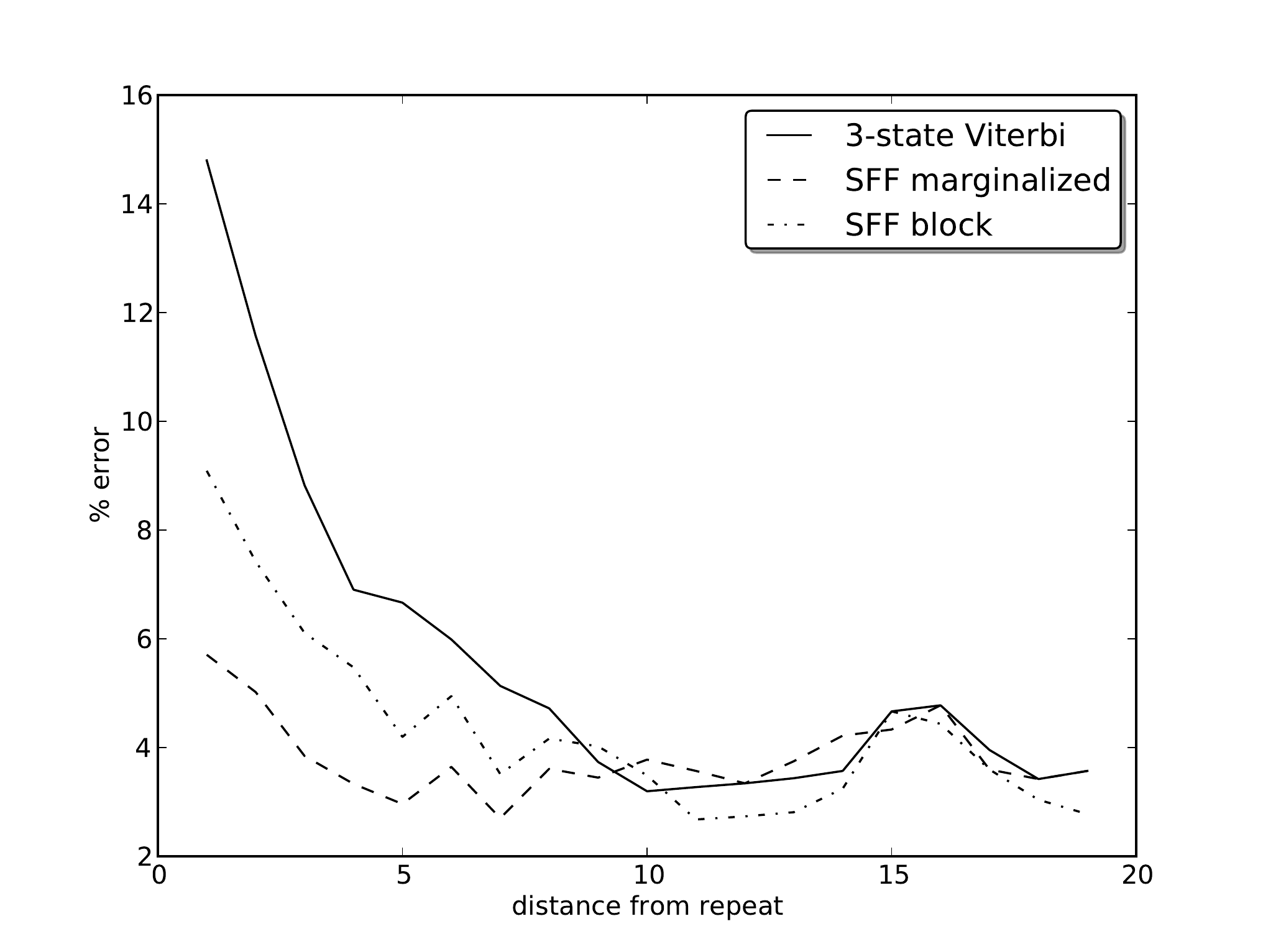}}
\caption{Alignment error rate of three decoding methods as a function of 
the distance from the nearest repeat.}\label{error_distance}
\end{figure}

We also compared the accuracy of predicting repeat block boundaries
(table columns block sensitivity and specificity). The number of
blocks with correctly predicted boundaries is quite low,
most likely because there are usually many high-probability
alternatives with slightly shifted boundaries. However, even though
more than half of the repeat blocks have some error in the boundary
placement, the SFF-based methods improve the alignment accuracy most markedly
close to repeat boundaries, as shown in Fig.\ref{error_distance}. 
This is expected, because far from repeats the
model works similarly to the three-state PHMM.

To illustrate the feasibility of running our methods on real genomic
data, we have realigned the 1.8Mb CFTR region on the human chromosome
7 to orthologous portions of the dog genome. We have started with the
lastz alignment \citep{Harris2007} 
downloaded from the Ensemble website \citep{Flicek2013}. We have then run
the TRF program on both species and selected alignment windows of
length 50--350 that contained at least one repeat. Regions without
repeats or with very long repeats were left with the original
alignment. Using the block decoding with the SFF model, we have thus
realigned windows covering roughly 70\% of the original alignment.
About 8\% of all alignment columns were annotated as repeats.




\section{Conclusion}

We have designed a new pair hidden Markov model for aligning sequences
with tandem repeats and explored a variety of decoding optimization
criteria for its use. The new model coupled with appropriate decoding
algorithm reduces the error rate on simulated data, especially around
boundaries of tandem repeats. With suitable heuristics, our approach
can be used to realign long genomic regions.

Our experiments are the first study comparing a variety of different
decoding criteria for PHMMs. Our criteria for the SFF model optimize
both the alignment and the repeat annotation. Depending on the
application, one or the other may be of greater interest, and thus
one may want to marginalize over all alignments and optimize the
annotation, as in \citep{Satija2010}, or marginalize over labels and
optimize the alignment.

Our model does not take into the account the dependencies between the
repeat occurrences in the two species. A tractable model allowing such
dependencies would be of great interest. Previously, we have explored
the problem of aligning two sequences simultaneously to a profile HMM,
but we were not able to design a simple generative model for this
purpose \citep{Kovac2012}.

\paragraph{Acknowledgements.} This research was funded by
VEGA grant 1/1085/12.

\bibliographystyle{apalike} \bibliography{main}

\end{document}